\documentclass[12pt,a4paper]{panl}
\usepackage{cite}
\usepackage{wrapfig}
\usepackage{graphicx}
\usepackage{amssymb}
\usepackage{amsfonts}
\usepackage{amsmath}
\usepackage{longtable}
\usepackage{rotating}
\usepackage{lscape}
\usepackage{epsfig}
\usepackage{multirow}
\usepackage{hyperref}
\originalTeX
\begin{document}

\title{Nearly forgotten results in development of physical cosmology}
\maketitle
\authors{A. F.\,Zakharov$^{a,b,}$\footnote{E-mail: alex.fed.zakharov@gmail.com}}
\setcounter{footnote}{0}
\from{$^{a}$\, National Research Center ''Kurchatov Institute'', Moscow Russia}
\from{$^{b}$\, Bogoliubov Laboratory for Theoretical Physics, JINR, Dubna}

\begin{abstract}
\
\vspace{0.2cm}

NICA is a tool to investigate the early Universe in laboratory. It would be reasonable to recall some essential issues in physical cosmology development. GR was created by A. Einstein in 1915. In 1917 Einstein proposed the first (static) cosmological model. Soon after the A. Eddington proved that the model is unstable therefore it can not be realizable in nature. In 1922 and 1924 A. A. Friedmann found non-stationary solutions for cosmological equations written in the framework of GR. In 1927 G. Lemaître obtained very similar results and, in addition, he derived the Hubble law (E. Hubble obtained this law from observations). Unfortunately, G. Lemaître published his paper in not very popular Belgium journal. In 1931  Lemaître proposed the first version of hot Universe model (he called it hypothesis of the primeval atom). In his book «The Primeval Atom – An Essay on Cosmogony» Lemaître predicted even a background radiation as a signature of his model. At the end of 1940s G. Gamow and his students proposed his hot universe model where he explained primordial nucleosythesis of helium at the initial stage of the Universe evolution. One of the properties of Lemaître – Gamow model was an existence of CMB radiation with a temperature around a few K. It was recalled that the discovery of CMB radiation was done by T. Shmaonov in 1956 
and his paper was published in 1957 (several years before the discovery done by Penzias and Wilson). In 1965, 1970 E. B. Gliner proposed vacuum like equation of matter which could correspond to exponential explosion of the Universe which was later called inflation. For decades in USSR, A. A. Friedmann’s cosmologocal non-stationary models were treated as purely mathematical results without cosmological and astronomical applications. On September 16, 1925 Friedmann passed away untimely and it would be reasonable to remind today his great contribution in physical cosmology since the authors book on Friedmann wrote “Similarly to Copernicus who forced the Earth to move, Friedmann forced the Universe to expand”.

\end{abstract}
\vspace*{6pt}

\small
\label{sec:intro}
\section{Introduction}

General Relativity (GR) was created after intensive conversations between A. Einstein and D. Hilbert in November 1915 \cite{Earman_78,Vizgin_01,Logunov_04}. 
Using his new theory of gravity  in November 1915   A. Einstein explained the problem of the anomaly of Mercury's motion, discovered in the middle of the 19th century by Le Verrier and remained unexplained for several decades. Einstein also considered the problem of deflecting a ray of light in a gravitational field, and it turned out that the angle of deviation in general relativity exceeds the angle of deviation in the Newtonian theory of gravity by two times \cite{Einstein_15} (see, English translation in \cite{Einstein_79}).

In 1917 A. Einstein found a static cosmological solution assuming that the Universe is homogeneous and isotropic \cite{Einstein_17} (see English translation in \cite{Einstein_52}). However, the assumptions adopted in the first cosmological paper were very useful and despite the fact that assumptions about the homogeinity and isotropy of the universe do not look very plausible, such assumptions turn out to be applicable to the theoretical description of astronomical observations. The assumptions  are still used in a large number of studies on cosmology and are often called the cosmological principle. 
To create a static cosmological solution and to compensate gravitational attraction A.~Einstein introduced so-called $\Lambda-$ term  which corresponds to repulsion.
In many textbooks and articles the authors repeated after G. Gamow that ''he [A. Einstein] remarked that the introduction of the cosmological term was the biggest blunder he ever made in his life'' \cite{Gamow_70} (see also a similar sentence in a popular article \cite{Gamow_56})). However, taking into account Gamow's personality, in particular, his predilection for jokes and practical jokes, and the different social status of Gamow and Einstein, it seems very strange that Einstein chose Gamow as an interlocutor to whom Einstein would admit to the biggest blunder of his entire life (see also discussions in \cite{Livio_11,Rosen_13}). Moreover, in 1998 it was discovered an accelerated expansion of the Universe and in order to explain this phenomenon, it is necessary to introduce a non-zero value of the lambda term  into the Einstein equations.
Otherwise, it is necessary to use a some generalization of the lambda term, which  has been called dark energy since 1999 \cite{Turner_99,Turner_99b}.

When discussing Einstein's static cosmological model, it should be noted that it is hardly applicable to describe the behavior of the universe because as a famous Russian mathematician V. I. Arnold repeated only stable mathematical solutions exist in physical world. Really, analyzing properties of the static cosmological model in 1930 A. Eddington  ''working in conjunction with Mr. G. C. McVittie'' proved that the Einstein's static cosmological solution is unstable \cite{Eddington_30}.

\label{sec:start}
\section{Start of GR studies in Russia}

In the early 20th century, despite significant achievements in some areas of physical science development in Russia, world leaders in physics worked in Europe.  Young Russians who wanted to acquire scientific skills at the cutting edge came to Europe for training and internship. One of these young people interested in science was a representative of the elite of the Russian empire (his father and grandfather were governors in Russian empire), Vsevolod Konstantinovich Frederiks (13.04.1885 (Warsaw) –-  06.01.1944  (Gorky)).  He played a key role in initiating research in Russia in the field of GR \cite{Vizgin_88,Sonin_88}. In 1903 Vsevolod Frederiks finished gymnasium in Nizhnij Novgorod and decided to go to Switzerland due to its convenient location to visit other European countries.
In 1907 he finished Geneva University with specialisation in physics and in  1909 he got PhD degree in physics under supervision of professor G. E. Guye (Geneve).
Since 1911 Frederiks was an assistant  under professor Woldemar Voigt\footnote{In 1887 W. Voigt found transformations which were similar to Lorentz ones \cite{Voigt_87}. In addition, Voigt  described group properties of his transformations as it was noted by H. Minkowski in his famous lecture delivered in 1908 \cite{Minkowski_52} (see also  discussions about a relation between Voigt and Lorentz transformations in
\cite{Miller_81,Pais_82,Ernst_01,Heras_14}).} at the Theoretical division of Physics Institute at G\"ottingen.
In 1914 the WWI began and Frederiks started to be a civil prisoner since he was a citizen of an enemy country. 
In 1910s famous mathematician David Hilbert was very interested in physics he was searching for an educated physicist to discuss actual physical issues. Frederiks began working as a Hilbert's personal assistant, and Hilbert paid Frederiks a monetary reward from his personal funds. The management of the University of G\"ottingen did not approve  this practice to contact with representative
of an enemy country in University and refused to allow Fredericks to enter the University in response to Hilbert's appeal.
In 1918  Frederiks came back in Russia and for a short period he worked in Institute of Physics and Biophysics in Moscow.
In 1919 he moved to Petrograd and was a senior physicist at State Optical Institute, a member of Atomic commission, an associate professor in Petrograd State University, a professor in Pedagogical Institute, in 1920 he was a lecturer at Polytechnic Institute. 
In 1921  Frederiks published the first Russian review on general relativity (GR) in  Soviet Physics Uspekhi which was  a  leading physical journal in Russia \cite{Frederiks_21}. 
In 1923 Frederiks was a senior physicist in  Institute of Physics and Technology, the Institute was founded by A. F. Ioffe in 1918 and in 1924 it was named the Leningrad Institute of Physics and Technology (LIPT) or shortly the Ioffe Institute.
V. K. Frederiks and A. A. Friedmann\footnote{British expert in astronomy and history of science  S. Mitton  noted \cite{Mitton_20}:  ''There are three ways to spell his surname: Fridman, Friedmann, and Friedman... Friedmann is the most common version in use today, largely in connection with cosmology. I [S. Mitton] use Friedman when referring to his publications because that is the version he used as author. A Bolshevik Government Decree issued 1917 December 23 eliminated double letters in Russian orthography. Typographers were required to enforce the reforms. The contemporary Russian literature invariably uses Friedman.'' I [AZ] basically use Friedmann in the paper since this spelling was used in a majority of articles and books that I read before writing the paper including English translation of the remarkable book \cite{Tropp_93} written by the best experts in history of Russian cosmology.} decided to wrote a joint book on foundations of GR but in 1924 it was published only the first chapter of the book. In the chapter only a tensor calculus was discussed (this project to write a joint book was not realized due to the untimely death of A. A. Friedmann).
 In Autumn 1927 Frederiks married  Maria Dmitrievna Shostakovich (a sister of composer D. D. Shostakovich). 
 In 1931 Frederiks was appointed as the Head of the Crystallization Laboratory in  LPTI 
 and in 1933  started to be the Head of anisotropic liquid Laboratory in LPTI.
 Frederiks studies in liquid crystals  were rather successful, in particular in 1927 V. Frederiks and A. Repiewa  discovered the phase transition in liquid crystals under action of magnetic field. Now this property of liquid crystals is used in operation LC screens in different electronic devices and it is called now the  Frederiks transition \cite{deGennes_93}.
 In 1934  Frederiks got DSc degree without a formal defense, he was nominated by the LPTI Scientific Council as a candidate for corresponding member of Soviet Academy of Science but he was not elected.  
 Due to very rapid development of contemporary physics in 1920s and 1930s it was necessary to write new textbooks and V. K. Frederiks was the
 co-editor with A. Р. Afanasiev of Course on General Physics (I. K. Kikoin, Yu. B. Khariton were among authors of chapters in the book). Later, the textbook was republished where only
  A. Р. Afanasiev was indicated since Frederiks was in detention at this time.
On October 20, 1936 V. K. Frederiks (among many other Leningrad scientists) was arrested   as a defendant in the Pulkovo case.
On May 23, 1937 Frederiks sentenced to 10 years in prison camps (Taishetlag). 
 On January 6, 1944  Frederiks died. Only in 1957 his relatives got an official document on his death,  where is the dash was written in the place of death. Lydia Chukovskaya (whose husband Matvey Bronstein\footnote{M. Bronstein was one of the leaders of Soviet theoretical physics \cite{Gorelik_94} and he was among the pioneers in  quantum gravity research \cite{Rovelli_02}. In the quoted paper C. Rovelli also noted that an introduction of the graviton concept proposed by D. I. Blokhintsev and F. M. Galperin in 1934 was an important initial step in development of quantum gravity.}  was an another Pulkovo case defendant) received a similar document  concerning a time and a place of the Bronstein death \cite{Chukovskaya_23}.

\section{A. A. Friedmann and dawn of physical cosmology}
\label{sec:Friedmann}

There's no doubt that Friedmann's cosmological work laid the foundations of physical cosmology and is a source of pride for Russian and international science \cite{Belenkiy_12,Soloviev_22}. For Friedmann's centenary, which was celebrated in June 1988, a scientific biography was written \cite{Tropp_88} (this remarkable book was translated into English in 1993).
In the book the authors wrote, ''He [Friedmann] discards a centuries-old tradition, which a priori, prior to all experience, considered the Universe eternal and forever motionless. He is making a real scientific revolution. Just as Copernicus forced the Earth to revolve around the Sun, so Friedmann forced the Universe to expand''.
In the days of the Baldin seminar we celebrated a memory of  this remarkable scientist. 
Alexander Alexandrovich Friedmann (Friedman) was born on 4(16).06.1888 in Sankt-Peterburg (SPb) and died  on 16.09.1925  in Leningrad (Sankt-Peterburg (1703 -- 1914), Petrograd (1914 -- 1924), Leningrad (1924 -- 1991),   Sankt-Peterburg (1991 -- till now)).
In 1897 - 1906   Alexander Friedmann was a student at the Second SPb gymnasium, in  1905 he wrote the first mathematical paper (published in 1907).
In 1906 - 1910 he was student at mathematical division of the faculty of physics and mathematics.
In 1910 - 1913 he left in the SPb University for a preparation for a professor position.
In 1914 – 1916  Friedmann joined the Russian army as a volunteer,   he served in aviation units. 
In 1918 – 1920 he was a professor of mechanics department of the Perm University. 
In 1920 – 1924 he was a researcher at the Atomic Commission in the State Optical Institute in Petrograd.
In 1920 – 1924 he was a professor at the Faculty of Physics and Mechanics of the Petrograd Polytechnic Institute. 
In 1920 – 1925 Friedmann  was a senior physicist, head of mathematical bureau, scientific secretary and since February 1925 director of the Main Geophysical Observatory.
In 1922  Friedmann published his first cosmological paper \cite{Friedmann_22} (English translation of the paper was published in \cite{Friedmann_99}).
In this paper, Friedman showed a solution describing an expanding universe that satisfies Einstein's equations.
Einstein read the Friedmann's paper and he published a short comment \cite{Einstein_22} (consisting of a few sentences and one expression) where he claimed:''The results of the work on the non-stationary world contained in the mentioned work seem suspicious to me.  In fact, it turns out that the solution indicated in it does not satisfy the field equations.''\footnote{A Russian translation of the comment published in the special issue of Soviet Physics -- Uspekhi dedicated to the 75th anniversary of Friedmann \url{https://ufn.ru/ufn63/ufn63_7/Russian/r637g.pdf}.}
Friedmann was extremely disappointed by this opinion of the GR creator. Friedmann consulted with Petrograd mathematicians and sent a lengthy letter to Einstein where Friedmann presented formulas confirming his conclusions. At the end of his letter Friedmann requested Einstein \cite{Tropp_93}
''...Should
you find the calculations presented in my letter correct, please be so kind
as to inform the editors of the Zeitschrift fur Physik about it; perhaps in
this case you will publish a correction to your statement or provide an
opportunity for a portion of this letter to be printed.''
Friedman did not receive Einstein's reply to this letter, possibly because Einstein was on a long overseas voyage at the time.
In 1923,  Petrograd theorist Yuri Krutkov was going to Germany on a scientific visit. Friedmann discussed the essence of his arguments with Krutkov in the letter to Einstein and requested Krutkov to meet with Einstein and convince him that Friedmann was right. In Leiden Krutkov met Einstein at the Paul Ehrenfest house (both Krutkov and Einstein were Ehrenfest's good friends) and Krutkov convinced Einstein that Friedmann was right. After that Einstein
wrote a short note where he mentioned that ''My criticism, as I saw from Friedmann's letter, communicated
to me by Mr. Krutkov, was based on an error in calculations. I consider
Friedman's results to be correct and shed new light...'' \cite{Einstein_23}\footnote{Russian translation of the Einstein note published in \url{https://ufn.ru/ufn63/ufn63_7/Russian/r637h.pdf}.}
In 1923  Friedmann published his book “World as space and time” (it was the first popular presentation of GR in Russia). The cover page of the book is presented in Fig. 1.
\begin{figure}[th]
\begin{center}
\includegraphics[width=50mm]{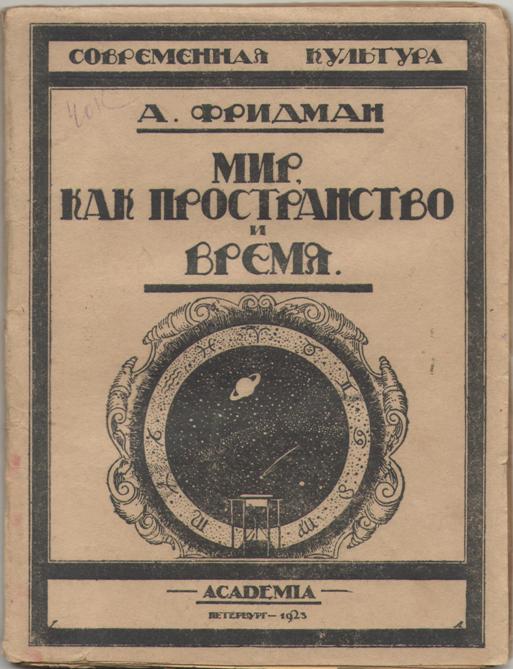}
\vspace{-3mm}
\caption{Cover page of the first Russian book on GR.}
\end{center}
\label{fig:book23}
\vspace{-5mm}
\end{figure}

In 1924 Friedmann   published his second cosmological paper, where the negative curvature case for the cosmological solution was considered \cite{Friedmann_24}. English traslation of the
paper was published in General Relativity and Gravitation \cite{Friedmann_99b}.\footnote{According to Fock's memoirs, he suggested  Friedmann to consider this case when Fock translated the first cosmological Friedmann's paper into German.}
Friedmann and Frederiks planned to write voluminous book ''Foundations of relativity theory'' for Russian readers but the authors wrote
only the first chapter ''Tensor calculus'' which was published in 1924.  The project was not finished due to Friedmann's untimely death.
Investigations of atmosphere turbulence were among main  Friedmann's interests and in July 1925 he and P. F. Fedoseenko\footnote{Pavel Fedoseenko
was the Osoaviakhim-1 captain and died in the balloon crash on January 30, 1934.}
 made a record-breaking balloon ascent (flight at 7400 m)
where the flyers did  medical experiments and measurements of atmosphere properties at different heights. After the flight Fedoseenko and Friedmann wrote their reports on their measurements.
In July – August 1925 Alexander Friedmann  and his wife (N. E. Malinina -- Friedmann) relaxed at the Crimea cost.
On August 17, 1925  Alexander Friedmann  came back to Leningrad, while his wife  went to another town where she was supposed to work at the geomagnetic station.
Perhaps, in his last train trip  Friedmann was infected by a typhus since he ate dirty fruits.   According to Friedmann himself, he probably got infected by eating an unwashed pear bought at one of the railway stations on the way from Crimea to Leningrad.  Suddenly (on September 2) he felt sick.
On  September 16, 1925  Friedmann  died in hospital (in days of the Baldin Seminar scientific community
the gravitational community remembered this remarkable man and his works on cosmology, in which it was firstly discovered that the universe can expand).
He was buried at the Smolensk Orthodox Cemetery in Leningrad (his tomb is presented in Fig. 2).
As we discuss below models of evolving Universe were practically banned in USSR in 1933 -- 1963 since the birth of the Universe may be easily
associated with the Genesis in Bible text (we will discuss the issue below in the Chapter on Lemaître studies).
In connection with the 75th anniversary of the birth of A. Friedmann, the Soviet authorities lifted the tacit ban on mentioning Friedmann's cosmological works, and in June 1963, the physical-mathematical division of the Academy of Sciences organized a special session dedicated to his memory.
At his introductory speech  P. L. Kapitsa  said at the  session  in June 1963 \cite{Kapitsa_74}:''Friedmann’s
name has so far been undeservedly forgotten. This is unfair and it needs to be fixed. We
must perpetuate this name. After all, Friedmann is one of the pioneers of Soviet physics, a
scientist who made a great contribution to domestic and world science.''
Soon after the session in July 1963  Soviet Physics Uspekhi published a special issue with articles of outstanding scientists
on subjects which were connected with Friedman's scientific interests.
In his short note about a genesis of GR studies in Russia
V.A. Fock wrote \cite{Fock_64}: "Alexander Alexandrovich Friedmann and Vsevolod Konstantinovich Frederiks were professors of Petrograd University (now Leningrad University) and they were the first scientists who started to teach Russian physicists (working in Petrograd) GR...Main speakers were V.K. Frederiks and A.A. Friedmann. Styles of their talks were rather different. Frederiks clearly understood physical aspects of theory and disliked to show mathematical calculations, while Friedmann placed emphasis on mathematics but not on physics. He aimed at mathematical rigor and paid attention to complete and precise formulations of initial assumptions. Discussions between Frederiks and Friedmann were very interesting."
At the session in 1963 it was decided to publish selected works of Friedmann as an academic publication edition \cite{Friedmann_66} (it was very convenient for Soviet researchers and students to read Russian translations of fundamental Friedmann's papers instead of difficult-to-access original articles written in German).
In June 1988 Soviet Academy of Sciences organized a big international Friedmann seminar in Leningrad to celebrate the 100th anniversary of his birth.
\begin{figure}[t]
\begin{center}
\includegraphics[width=80mm]{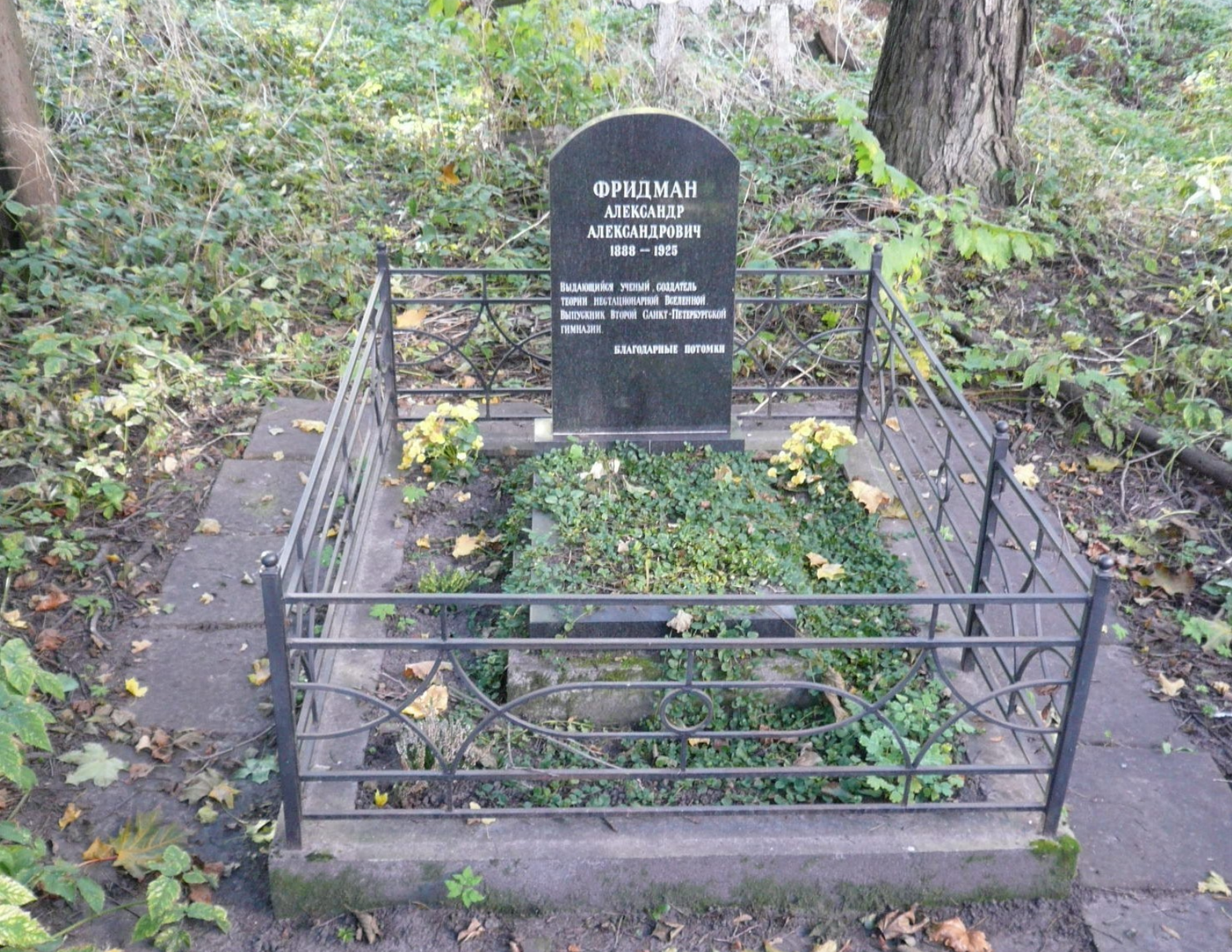}
\vspace{-3mm}
\caption{The Friedmann's tomb at the Smolensky Cemetery in St. Petersburg. The photo was taken in October 2025.}
\end{center}
\label{fig:tomb}
\vspace{-5mm}
\end{figure}

\label{sec:Lemaitre}
\section{G. Lemaître , the Hubble law and the Big Bang concept}
Georges Henri Joseph Édouard Lemaître (17 July 1894 – 20 June 1966) was a Belgian Catholic priest who made major contributions to cosmology.\footnote{
Lemaître was born in Charleroi, Belgium
approximately 20 year before WWI. In 1911, Lemaître started to study engineering at the Catholic University of Louvain. In 1914, after the outbreak of World War I, Lemaître interrupted his studies and he entered the Belgian army as   a volunteer.  After the war, Lemaître abandoned engineering for the study of physics and mathematics.
 In the 1920s,  Lemaître got scholarship and spent a few years in the USA, studied in Harvard and worked at the Harvard Observatory. In 1925 he came back to Belgium  became a part-time lecturer at the Catholic University of Louvain. 
After a defence of his PhD thesis in MIT 
in 1927 he got a ordinary professor position in Louvain. Lemaître  personally knew Vesto Slipher (for instance, it is known that he visited Slipher in Arizona and Hubble at Mount Wilson in the summer of 1925 \cite{Farrell_05}). Slipher analyzed Doppler spectral shifts and found that distant
nebulae are preferably moving from us \cite{Slipher_22}. E. Hubble used Slipher’s data on velocities of distant nebulae, and
now some historians of science proposed to rename the Hubble constant \cite{MacDougal_24}.
Interesting details on   Lemaître's life and scientific investigations are given in \cite{Lambert_15}.
}
G. Lemaître played an exclusive role in a development of cosmological studies. 
As a recognition of his achievements at the Symposium devoted
to 50 years since the introduction of the {\it Primeval Atom}  concept  an outstanding cosmologist James Peebles (he got  Nobel prize in physics in 2019) wrote \cite{Peebles_84}:
''Physical scientists have a healthy attitude toward the
history of their subject: by and large we ignore it. But it is
good to pause now and then and consider the careers of those who
through a combination of the right talent at the propitious time
have had an exceptional influence on the progress of science.
As I have noted on several occasions it seems to me that Georges  Lemaître
played a unique and remarkable role in setting out the
program of research we now call physical cosmology''.
The results similar to Friedmann's were independently obtained by the Belgian Abbe George Lemaître.
In addition Lemaître  derived of the law which describes velocities of distant objects as dependence on
distances toward them was observationally discovered by E. Hubble, and later this law began to
bear his name. While Friedmann was a skillful mathematician and he did not speculate about astronomical applications of his results, Lemaître
had a deep knowledge in astronomy and always think about an opportunity to test theoretical laws with astronomical observations.
So, in 1927 Lemaître published paper  \cite{Lemaitre_27}  in an obscure Belgium journal where the expanding Universe solution of the Einstein equations were found and he analytically derived
a relation between distance and velocity  of distant galaxies (now it is called Hubble -- Lemaître law). He obtained  about 575 km/sec/Mpc \cite{Lemaitre_13,Luminet_13}, not very  different from Hubble’s ultimate value. 
 From 24 to 29  October 1927, the 5th Solvay Congress on physics in Brussels was organized and Einstein  participated at this activity. The main topic was ''Electrons and Photons''  
 or in other words,
 discussion of main quantum mechanics laws which were discovered in 1925. 
  H. Lorentz was the chair of the Congress (he was also the chair at the four previous  Solvay Congresses on physics).
 Lemaître was not invited to those discussions since he was not so famous  as other participants of the Congress,\footnote{Participants of the Vth Solvay Congress in physics are presented
   in \url{https://en.wikipedia.org/wiki/Solvay_Conference\#/media/File:Solvay_conference_1927.jpg}. 
   As the History Insider noted it is ''The Most Intelligent Photo Ever Taken''
   since  17 Nobel Prize Winners are presented at this photo.}
   however, Louvain is only 20 km from the Belgium capital. Lemaître went to 
  the Congress venue and during the Congress breaks Lemaître had conversations with Einstein about his cosmological solutions published in 1927.
  Einstein commented favourably the Lemaître's mathematical skills but he rejected an expanding Universe concept. 
  ''Your calculations are correct, but your physical
insight is abominable,'' concluded Einstein  \cite{Bartusiak_09}.
  Einstein also directed  Lemaître's attention at the Friedmann's cosmological studies.
  In spite of Einstein's skepticism in respect to models of expanding Universe, Lemaître continued to work on interpretation since he knew Slipher's results that majority of distant galaxies
  have redshifts \cite{Bartusiak_09,Mitton_17}, it means that most galaxies are moving away from us.
  Using Cepheid variable stars (who are giant stars where there is dependence between their luminosities and periods)  E.   Hubble found a linear dependence between velocities of extragalactic sources and distances toward them \cite{Hubble_29}, or in other words, $V=HR$, where $V$ is a velocity of extragalactic source, $R$ is a distance toward a chosen galaxy, $H$ is a constant, which is called now the Hubble constant (in this paper Hubble estimated the constant as 500~(km/s)/Mpc.\footnote{It should be noted that the law of the linear dependence of velocity (or redshift) on distance was discovered by Hubble based on the motion of nearby galaxies that are not uniformly and isotropically located, as Lemaître assumed after Einstein, and before Hubble, he obtained an estimate of the proportionality parameter in this linear relationship, which is close in value to Hubble's estimates.   At the IAU congress in 2018 in Vienna it was proposed and accepted to call this dependence as the Hubble--Lemaître law.  Many years after the discovery of the law J. Peebles wrote ''This nearly homogeneous expansion is in striking contrast to the extremely clumpy space distribution'' \cite{Peebles_89}.} Currently, astronomers estimate this constant and get a value about an order of magnitude smaller, which is very unusual for physical constants, and the main problem and inaccuracies are related to measuring distances.  
 
 As it was noted earlier, in 1930 A. Eddington proved that the static cosmological solution proposed by Einstein is unstable and at the Meeting of Royal Astronomical Society on January 10,
 de Sitter expressed doubts in respect to the Einstein's model of static Universe taking into account Hubble's results and at Meeting it was declared that it is necessary to search for non-stationary cosmological solutions.  These issues  were published in February Issue of  Observatory Journal. Usually Lemaître read the journal (in particular, this February issue)
 and
 sent a letter to Eddington \cite{Lambert_15} where he informed that he solved the problem 3 years ago: ''Dear Professor Eddington,
I just read the February n° of the Observatory and your suggestion
of investigating a nonstatistical intermediary solution between
those of Einstein and de Sitter. I made these investigations two years
ago. I consider a universe of curvature constant in space but increasing
with time. And I emphasize the existence of a solution in which
the motion of the nebulae is always a receding one...''
Lemaître also attached copies of his paper \cite{Lemaitre_27} to the letter to Eddington.
After reading the Lemaître's letter Eddington understood that the problem to find a suitable cosmological solution which is consistent with astronomical observations has been solved
and he started immediately to promote this breakthrough  \cite{Mitton_20}.  In particular, in a review \cite{Eddington_30b} on a Silberstein’s book   Eddington wrote
''Three years ago a very substantial advance in this
subject was made by Abbé G. Lemaître...it renders obsolete the contest between Einstein’s and
de Sitter’s cosmogonies...  We can now prove that Einstein's universe is unstable. The equilibrium having been disturbed, the universe will progress through a continuous series of intermediate states towards the limit represented by de Sitter’s universe. By Lemaître’s analysis the history of this progress can be studied; and the intermediate stages (one of which must represent the present state of the world) can be treated in detail.''
Eddington was sure that is was found the solution that described a behavior of the Universe and he wrote \cite{Eddington_31}: ''We have recently learnt mainly through the work of Prof.  Lemaître that this spherical space is expanding rather rapidly.''
Eddington recommended to  publish an English translation of Lemaître's paper \cite{Lemaitre_27} in Monthly Notices of the Royal Astronomical
Society (MNRAS) and after that William Smart, the Editor of MNRAS and RAS Secretary sent a letter to Lemaître requesting him to submit
an English translation of his 1927 paper and to propose to him to become
a member of this Society \cite{Lambert_15}. In Smart's letter it was written that the translation should have not more than 72 paragraphs, and  following Smart's recommendations Lemaître
reduced an article length and 
deleted the piece of text where the quantity (which called now the Hubble constant) was evaluated
from available observational data on  redshifts of moving galaxies. 
Some authors declared that it was a free Lemaître's choice to do not indicate in 1931 that in 1927 Lemaître discovered the law now named after Hubble  \cite{Livio_11b} (since 2018 
it is called the Hubble -- Lemaître law). However, other authors
 claimed that this decision was made under pressure from Smart  \cite{Mitton_20} and for English speaking people Hubble had the absolute priority in discovering this law.
In 1931 Lemaître translated  his paper written in French \cite{Lemaitre_27}  and published the English translation in MNRAS
\cite{Lemaitre_31}. Also in 1931 Lemaître considered the birth of the Universe as a quantum physics  phenomenon \cite{Lemaitre_31b}. It was the first introduction of a hot cosmological model
(which were called Big Bang following  Hoyle's ironical cliche). Later, these Lemaître's ideas were transformed in the Primeval Atom hypothesis. 
It would interesting to note that Lemaître predicted a background radiation with a temperature around a few kelvins in his cosmological model of hot Universe. 
more precisely, in 1934 Lemaître wrote \cite{Lemaitre_34} ''...if all the atoms of the stars were equally
distributed through space there would be about one atom per cubic yard,
or the total energy would be that of an equilibrium radiation at the temperature
of liquid hydrogen..'' (see also comments in \cite{Luminet_11}).
Lemaître hoped that his cosmological approach may explain an origin of cosmic rays (at this time Robert Millikan was interested in the subject and he proposed cosmic rays as a term for high enrgy particles from space). 
At the end of 1932,  Millikan invited Lemaître to visit Caltech. On January 11, 1933,
Lemaître was invited to deliver a lecture at the Mount Wilson Observatory, where E. Hubble
worked. A. Einstein followed this Lemaître’s lecture. When journalists asked Einstein about
his impression of Lemaître’s cosmological model, Einstein replied, “This is the most beautiful
and satisfactory explanation of creation to which I have ever listened!”
\footnote{D. Lambert, Einstein and Lemaître : Two friends, two cosmologies..., 
\url{https://inters.org/einstein-lemaitre}.
}. 
After that journalist Dunkan Aikman published interview with Lemaître with the title ''Lemaître follows two paths to truth: The famous scientist, who is also a priest, tells why he finds no conflict between science and religion'' (the interview was published in New York Times on February 19, 1933
\footnote{\url{https://inters.org/files/lemaitre_two_paths_truth_nyt1933.pdf}.}. This article was extremely popular over the world.
Because the theses that religion can open the way to truth, and that religion does not contradict science, were unacceptable to Soviet ideologists. They found a way out of this situation, just as it was later done for genetics and cybernetics. Soviet ideologists came to the conclusion that the approach to constructing cosmological models based on general relativity should be declared pseudoscientific, and the Soviet scientists who worked on this topic should be declared pseudoscientists and admirers of Western bourgeois science. Thus, a division occurred in cosmology: Western cosmology, which considered realistic dynamic models of the Universe, and Soviet cosmology, which asserted that the Universe is infinite in time and space.
It was done in contradiction to a famous Chekhov's statement that there is no national science as there is no national multiplication table. So,  Soviet ideologists declared that the Universe must be infinite in space and in time in contrast Friedmann --  Lemaître point of view on the subject.

Even if famous Soviet scientists considered non-stationary cosmological models describing expanding Universe in this case, they were rather strongly criticized by representatives of the Communist Party apparatus, and, unfortunately, some of the critics were quite good professionals in mathematics and physics and in spite of that they participated in actions on pressure on genuine science and its representatives. 
Let us consider one interesting case when Soviet scientific community decided 
to celebrate half-century since the creation  of special relativity and the Division of Physics and Mathematics decided to organize a Session devoted to this event on November 30 -- December 1, 1955.
 After the Session, on January 16, 1956 instructor of the Central
Committee of the Communist Party of the Soviet Union (CC CPSU) A. S. Monin\footnote{\url{https://en.wikipedia.org/wiki/Andrei_Monin}.} and the
head of the Science Department of CC CPSU V. A. Kirillin\footnote{\url{ https://en.wikipedia.org/wiki/Vladimir_Kirillin}.}  sent a letter to the Presidium of
the Soviet Academy of Sciences where they strongly criticized the Session where achievements
of relativistic cosmology were discussed. The authors of the letter declared that the talks of
E. M. Lifshits, Ya. B. Zeldovich, L. D. Landau and V. L. Ginzburg had significant disadvantages
and were not criticized properly by the audience \cite{Blokh_05}. At the beginning of the letter, it was
written, ''...The program of the session, prepared by the Organizing Committee, which included
Academicians Tamm and Landau, Corresponding Member of the USSR Academy of Sciences
Ginzburg and Professor Lifshitz, was unsatisfactory, as it included reports by Academician
Landau, Corresponding Member of the USSR Academy of Sciences Ginzburg and Professor
Lifshitz, who do not work in the field of relativity theory and are known for their nihilistic
attitude to the development of methodological issues of this theory''. Time has passed, and
now we know the importance of the criticized scientists for Soviet theoretical physics, and it
is hard to believe that their colleagues criticized them  genuinely, since both Monin and Kirillin later
became great Soviet organizers of science and full academicians in the field of physics and its
applications. 
Concerning E.~M.~Lifshits report at the session the letter authors wrote
''This report is devoted to a review of research on relativistic cosmology. 
He was a significant advocate of the idealistic "theory of the expanding universe."  {\it This "theory", which is an illegal extension of non-stationary solutions of Einstein's gravitational equations to the universe as a whole, was built by Abbé Lemaître on the direct order of the Pope} [the italic font in this phrase was chosen by AZ]. According to this "theory," the universe has a finite age; at the time of its formation, it occupied an infinitesimally small volume, and then began to expand; such an expansion is taking place at the present time.''
In this letter the authors also criticized Tamm, who chaired the session where Lifshitz's report was heard. Thus, the authors criticized and labeled as unprofessional almost all Soviet theoretical physicists who later became Nobel laureates, namely, I.E.~Tamm, L.D.~Landau, V.L.~Ginzburg and only I.M.~Frank was not criticised (probably because he was not presented at the Session). Naturally, the question arises as to the professionalism and impartiality of the experts rendering such a verdict on the incompetence of the most prominent Soviet theorists.
  

\label{sec:Shmaonov}
\section{T. A. Shmaonov and the CMB discovery in Pulkovo}

Since 1948 G. Gamow started to work American project to create a superbomb or in other words, H-bomb.
During this period, refined cross-section data on the nuclear interaction of light elements became known, and these data could be used to calculate helium nucleosynthesis in the initial minutes of the evolution of the universe. 
At the end of  1940s, G. Gamow proposed his version of the hot Universe model where he estimated
a helium nucleosynthesis in primordial Universe \cite{Gamow_46,Gamow_48}, and soon after that his students calculated
Cosmic Microwave Background (CMB) radiation   \cite{Alpher_48} which should have now temperature
around 5 K. In consequent calculations the CMB temperature was slightly different, but always
it was a few kelvins. It would be reasonable to note that similar estimates for background radion temperature were done by Lemaître as we noted earlier.
The CMB (cosmic microwave background) 
radiation or relic radiation  in Russian literature  is one of the key signatures of the hot Universe model developed by G. Gamow
in the 1940s and 1950s. The CMB radiation was discovered by T. Shmaonov  at the Pulkovo Observatory several
years before A. Penzias and R. Wilson (who were awarded the Nobel prize in 1978) \cite{Shmaonov_57}.\footnote{In 1940, Andrew McKellar, analyzing the spectra CN and CH, found that interstellar medium was very
cold with a temperature of approximately 2 K, or more precisely 2.7, 2.1 and 0.8 K \cite{McKellar_40}. However, astronomers
did not have a correct explanation for this phenomenon for decades.}
Now Shmaonov's discovery is widely known not only in Russia but also in the world, for
instance, there is a description of this Shmaonov’s result in a fundamental book \cite{Peebles_09} 
(10 years after publication of this book Peebles was awarded by Nobel prize in physics for CMB studies), and see
also  Trimble's review \cite{Trimble_06}. 
 T. A. Shmaonov presented his lecture on the discovery 60 years after the events, see Fig.~3.
In Pulkovo observatory Shmaonov’s supervisors were professor Naum Lvovich Kaidanovsky and
professor Semyon Emmanuilovich Khaikin. 
Khaikin was one of the first
L. I. Mandelstamm’s students in Moscow, a great expert in theoretical physics, non-linear physics, radiophysics,
radio astronomy, and the dean of the Physics Department at Moscow State University. Khaikin
was fired from Moscow State University, MEPhI, and Lebedev Physical Institute after being
accused of promoting idealism and Machism. A nice book was recently published about this
remarkable scientist and teacher \cite{Yakuta_21}. When Shmaonov obtained his observational results he asked Khaikin about possible interpretation of his
discovery, and Khaikin replied that he had no explanation, but he recommended to published them because
it may be very important  \cite{Yakuta_21}. Since 1930s until 1960s Soviet physicists did not cite Gamow’s papers at
all, Khaikin worked as the head of the Radioastronomy Department at the Main Astronomical
Observatory in Pulkovo, Gamow was one of the brightest representatives of Leningrad school of
physics, and there is no doubt Gamow’s works were known to many leading Soviet physicists, such as
Khaikin, but he preferred not to declare that his PhD student (Shmaonov) had received a confirmation
of the Gamow’s predictions. 
Moreover, as indicated in the article on cosmology published in the Great Soviet Encyclopedia \cite{Zelmanov_55}, scientists discussing models of the expanding universe are recognized as supporters of bourgeois and religious philosophy, which is hostile to the Soviet one.
Earlier in the 1940s and 1950s, Khaikin had already been harshly criticized by his colleagues for allegedly promoting idealism and Machism at Moscow University and the Lebedev Physical Institute, which subsequently led to his dismissal. 
After defending his PhD thesis, T. Shmaonov left astronomy and later worked at the Institute
of General Physics in the laser physics laboratory headed by A. M. Prokhorov, who got the
Nobel prize in 1965 for the laser discovery and was Academician-Secretary of the Department
of General Physics and Astronomy (1973–1993). In 1978, when A. Penzias and R. Wilson got
their Nobel prize for CMB discovery, A. M. Prokhorov criticized Shmaonov that he did not
inform properly scientific community about his CMB studies and did not promote his discovery.
Sometimes people say that Shmaonov’s achievements were not known for many years, since
they did not have an appropriate cosmological interpretation. There is a popular opinion that
no one in the Soviet Union knew about the Gamow model of the hot Universe and the predictions
of this model. However, it seams to me this interpretation is not correct. 
After all, generally speaking, not discussing something publicly does not mean that the subject of discussion is not known.
The fact that the predictions of the hot universe model were not publicly discussed does not mean that these predictions were not known. So already in an article in 1963 in the jubilee collection dedicated to the 75th anniversary of Friedmann's birth, Zeldovich critically discussed these predictions done by G. Gamow.
As mentioned earlier,
in the Soviet Union the  models of expanding Universe (including Gamow’s ones) were considered
inadequate descriptions of the Universe and their consideration was not welcomed by
official ideology and philosophy. In addition, despite the fact that Gamow was one of the most
famous Soviet theoretical physicists, he did not return from a scientific trip abroad without the
permission of the authorities. Thus, the mention of Gamow’s works could be interpreted as a
support for his disloyal attitude towards the Soviet government.
A. D. Chernin called the story of the lack of a correct interpretation of Shmaonov's results a sad episode and noted that an American historian of science had found and interpreted Shmaonov's work in a cosmological context \cite{Chernin_94}.
In another article dedicated to Gamow's 90th birthday \cite{Chernin_94b}, Chernin showed how Gamow simply estimated the temperature of the cosmic microwave background radiation without a detailed analysis of primary nucleosynthesis.    
\begin{figure}[th]
\begin{center}
\includegraphics[width=50mm]{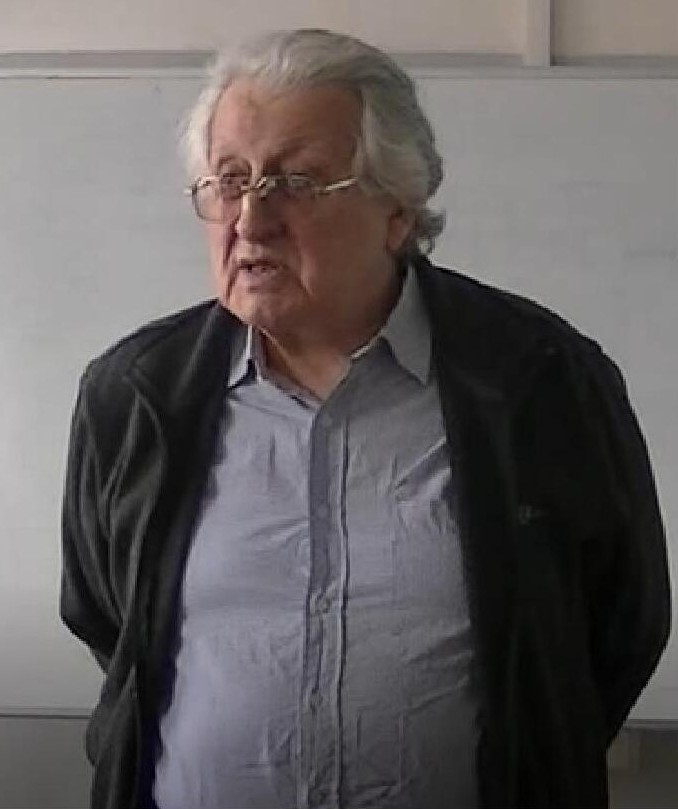}
\vspace{-3mm}
\caption{Tigran Aramovich Shmaonov presented his talk on CMB discovery in Pulkovo Observatory in 1956 -- 1957. The talk was given on 17 April 2017 at the S.I. Vavilov Institute for the History of Science (Moscow).}
\end{center}
\label{fig:Shmaonov} 
\vspace{-5mm}
\end{figure}

\label{sec:Relikt}
\section{Relikt-1 and COBE discoveries}
Because the Earth (together with Solar system, Galaxy, Local Group of galaxies) moves in respect to the CMB, a
dipole temperature anisotropy of the level of is expected.
A dipole temperature anisotropy of the level of $\Delta T/T = 10^{-3}$ was observed, it corresponds to a peculiar velocity
380 km/s of the Earth towards the constellation Virgo
(first measurements of dipole anisotropy were not very
precise \cite{Conklin_69}, but later the accuracy was significantly improved \cite{Smoot_77}. Moreover, based on results of balloon
measurements it was claimed several times about
features of quadrupole anisotropy, however, the
quadrupole term was too high in these experiments as further
studies had indicated, or reversely the realistic level
of quadrupole anisotropy was to low to be detected in the
balloon experiments.

In 1983, in the Soviet Union the Relikt-1\footnote{In Soviet Union the CMB radiation was called the relic (in Russian it sounds like Relikt) radiation.} experiment
was conducted aboard the Prognoz-9 spacecraft in order 
to investigate CMB radiation from space for the first time
in history. As many other Prognoz missions, the scientific
payload was prepared by the Space Research Institute of
the Soviet Academy of Sciences. Igor Strukov was
the principal investigator of the Relikt-1 project.
The spacecraft Prognoz-9, had an 8 mm band radiometer
with an extremely high sensitivity of 35 mK per second
and it was launched into a high apogee orbit with a
400,000 km semi-axis. The high orbit was a great advantage
of the mission since it allows to reduce of geomagnetic
field impact on measurements. A disadvantage of
the experiment was that the observations were conducted
only in one spectral band, therefore, it was a freedom
for a theoretical interpretation of the results, in contrast,
multi-band measurements provide very small room for
alternative explanations of anisotropy.
The radiometer scanned the entire celestial sphere for
six months. Computer facilities (and therefore, data analyzing)
were relatively weak in the time in Russia. Preliminary
analysis of anisotropy studies indicate upper limits on
anisotropy \cite{Klypin_87}, but in this case even the negative results
(upper limits) were extremely important to evaluate
a sensitivity to design detectors for next missions, including
COBE (COsmic Background Explorer).

At the NASA cite \url{https://lambda.gsfc.nasa.gov/product/relikt/}  it is given a brief information about Soviet Relikt-1 mission:
''Prognoz 9, launched on 1 July 1983 into a high-apogee (700,000 km) orbit, included the Relikt-1 experiment to investigate the anisotropy of the CMB at 37 GHz, using a Dicke-type modulation radiometer. During 1983 and 1984 some 15 million individual measurements were made (with 10\% near the galactic plane providing some 5000 measurements per point). The entire sky was observed in 6 months. The angular resolution was 5.5 degrees, with a temperature resolution of 0.6 mK. The galactic microwave flux was measured and the CMB dipole observed. A quadrupole moment was found between 17 and 95 microkelvin rms, with 90\% confidence level. A map of most of the sky at 37 GHz is available.'' 
After that at this cite references at key Relikt-1 publications are given where information about the Relikt-1 discovery was presented \cite{Strukov_92,Strukov_92b}.

In 1986, at the Space Research Institute it was decided
to prepare a next generation of space experiments
to study the anisotropy of CMB, and start to develop
the Relikt-2 project. The sensitivity of the detectors was
planned to be in 20 times better than the Relikt-1 sensitivity.
The Libris satellite was scheduled to carry the
Relikt-2 payload and the spacecraft was planned to be
located near the Lagrangian point L2 (in the Sun – Earth
system). 
Now  several spacecrafts, including JWST, are operating around the L2 point.
Originally, it was a plan to launch the Libris
spacecraft in 1993–1994, however, the project has not
been realized, basically due to a lack of funds. To prepare
Relikt-2, the team members re-analyzed Relikt-1
data and finally in beginning of 1992 they discovered
signatures of the quadrupole anisotropy, but I. Strukov
required to check the conclusions again and again. The
discovery of anisotropy by the Relikt-1 spacecraft was
first reported officially in January 1992 at the Moscow
astrophysical seminar and the Relikt-1 team submitted
papers in Soviet Astronomy Letters \cite{Strukov_92} and Monthly
Notices of Royal Astronomical Society \cite{Strukov_92b} (soon after
the papers were published).

The discovery of anisotropy by the Relikt-1 spacecraft
was first reported officially in January 1992 at the
Moscow astrophysical seminar. Relikt-1 team submitted
their paper in Soviet Astronomy Letters and Monthly
Notices of Royal Astronomical Society on January 19,
1992 and on February 3, 1992, respectively.
On April 21, 1992, G. Smoot (the head of DMR experiment
aboard the COBE mission) and his co-authors
submitted a paper in Astrophysical Journal (Letters) \cite{Smoot_92}. COBE orbit was rather low (around 1000 km) and therefore, it was a hard problem to separate
a signal and a noise which was very strong at the low orbit.
On April 22, 1992, Smoot reported at a press conference
about the discovery of the CMB anisotropy with
the COBE satellite. After that mass media reported these
results as the main science success. In 1992, COBE results
about discovery of the CMB anisotropy were reported
elsewhere. However, no doubt, the COBE collaboration
knew results of Relikt-1 team and even quoted
their upper limit on the quadrupole anisotropy published
at the paper \cite{Klypin_87}. 
However, summarizing, one could say that since papers
of the Relikt-1 team were submitted on January 19,
1992 and February 3, 1992 in Soviet Astronomy and
Monthly Notices of Royal Astronomy Society respectively.
 COBE paper was submitted on April 21, 1992,
one would conclude that the discovery one quadrupole
anisotropy was done by the Relikt-1 team and published
in papers \cite{Strukov_92,Strukov_92b} and also independently and almost simultaneously the phenomenon was discovered by the COBE-DMR  team.
In 2006 J. Mather and G. Smoot got Nobel prizes in physics for СMB studies since the COBE-FIRAS group measured CMB spectrum while COBE-DMR group 
found the quadrupole anisotropy.
The results of the Relikt-1 group were criticized by the members of the COBE group, but subsequent analysis showed that the results of the Relikt-1 group were in agreement with the results of the WMAP group  
\cite{Skulachev_10}.

\label{sec:Gliner}
\section{E. B. Gliner's ideas on cosmic vacuum, inflation and accelerated expansion of the Universe}

Until the end of the 1990s, theorists and astronomers thought that Einstein's
$\Lambda$-term is very small and they adopted $\Lambda =  0$. 
In 1998 astronomers discovered accelerated expansion of the Universe and in this case it is necessary to include into consideration  $\Lambda-$term or so called dark energy.
However, we have to remember that in 1965 Erast  Borisovich Gliner considered cosmological
(inflationary) models with accelerating expansion \cite{Gliner_65,Gliner_02} (see also an essay on the scientific life
of this remarkable scientist and personality in \cite{Yakovlev_23}). Such Gliner’s models looked very exotic
and some outstanding Soviet cosmologists (including Ya. B. Zeldovich) criticized them; on the
other hand, V. L. Ginzburg and A. D. Sakharov tried to support Gliner and to promote his studies \cite{Yakovlev_23, Chernin_13}. For instance, recalling his work on cosmology in the 1960s, A. D. Sakharov wrote that 
''Gliner wrote about the same thing independently and with greater certainty in the same years'' \cite{Sakharov_19}. 
However, in spite of the  undoubted  importance of his studies, Gliner’s results were nearly forgotten for decades \cite{Yakovlev_23}.

\section{Conclusion}

This fall marks the 100th anniversary of the death of A. A. Friedmann and the 110th anniversary of the GR creation.  The above are some results in the development of gravity and cosmology that are not covered in great detail in the Russian literature, for example, when discussing the Big Bang model in Russian literature, Lemaître's contribution is often forgotten, and the discovery of Shmaonov and the results of the Relikt-1 mission related to the discovery of the quadrupole anisotropy of the universe are very rarely mentioned.
Unfortunately, very often our compatriots are not active to promote achievements of other researchers (including our domestic achievements). 
In these circumstances people remember the M. I. Monastyrsky's dictum \cite{Monastyrsky_09} ''...By the way, we (in Russia) are very fond of lamenting the underestimation of Russian scientists in the West. Russian  (Soviet) scientists are the ones who hinder the recognition of Russian (Soviet) scientists the most, however, a careful analysis of the real facts shows...''
The discussions of the fact that the quadrupole anisotropy of CMB was detected simultaneously from the analysis of Relikt-1 and COBE data were presented in \cite{Zakharov_08_1,Zakharov_08_2}.
Some additional information related to the subject of this article can be found in historical introductions in the recent works \cite{Zakharov_25_1,Zakharov_25_2}.

\section*{Acknowledgements}

The author thanks the organizers of the XXVIth  International Seminar on  High Energy Physics Problems "Relativistic Nuclear Physics and Quantum Chromodynamics" (JINR, Dubna) for the invitation to present my contribution as an invited talk at this event and their attention to the research presented in the paper.
The author appreciates A. A. Baldin, \framebox{A. D. Chernin}, E. E.  Donets,  A. D. Kaminker, V. O. Soloviev, M. V. Tokarev, D. Ya. Yakovlev for useful discussions of the subject.
The author is grateful to the anonymous referee for constructive comments that helped to improve the text of the article.

\section*{Funding}
 This work was supported by ongoing institutional funding.
No additional grants to carry out or direct this particular
research were obtained.

\section*{Conflict of interest}
The author declares that he has no conflicts of interest.


\end{document}